# Spontaneous emergence of a spin state for an emitter in a time-varying medium.

Samuel Bernard-Bernardet[1], Marc Fleury[2], Emmanuel Fort[1]*

[1]Institut Langevin, ESPCI Paris, Université PSL, CNRS, 1 rue Jussieu, 75005, Paris, France.

[2] Two Prime, Open Finance Group, Atlanta, USA

*Correspondence to emmanuel.fort@espci.fr

## Abstract

Time varying media can dramatically modify the emission of embedded sources by producing time reversed waves refocusing on the source. Here we show that such a back action can create an angular momentum by setting the source in a spontaneous spin state. We experimentally implement this coupling using self-propelled bouncing droplets sources coupled to the surface waves they emit on a parametrically excited bath. The spin state dynamics result from a self-organized interplay between the source motion and the time reversed waves. The discrete stability analysis agrees with the experimental observations. In addition, we show that these spin states provide a unique opportunity for an experimental access to parameters enabling comparison and calibration of the various existing models.

## 1 Introduction

Time varying media offer new exciting possibilities to control and manipulate wave propagation such as parametric amplification [1], isolation and nonreciprocity [2–4] or frequency conversion [5,6]. One fascinating property of time varying media is the creation of time reversed waves which refocus on the emitting source by reflection on time interfaces [5,7–9]. This strongly enhances the back-action of the waves on the source by introducing a coherent feedback extended in time through the multiple temporal reflections [10,11]. The emission of a source is thus significantly modified when embedded in a time varying medium. When strong enough the feedback can break the symmetry of the emission and set the source into motion [12–14]. Here, we show that this coherent feedback can even self-organize to create a stable spontaneous spin state [15]. We implement the emergence of this angular momentum using self-propelled droplets, called walkers, bouncing on a vertically shaken liquid bath. These droplets can be considered as a generic model system of a point source interacting with its own emitted wave field in a parametrically excited medium. At each bounce the droplet emits waves. The vertical modulation of the bath induces a modulation of the wave speed generating time reversed waves refocusing on the droplet source. This coherent feedback destabilizes the droplet that becomes self-propelled [16–19]. The wave field driving the droplet is a superposition of the elementary waves produced at each bounce and sustained for a tunable time by the parametric excitation of the bath [18,19].

The existence of several dynamic stable states in 1D have already been shown [20–24]. The possibility of spin state was investigated theoretically by Oza et al. [25,26] inspired by experiments of walkers placed in a rotating frame [19,27]. Spin states refer to circular trajectories followed by walkers with discrete radii for which the wave field created by the previous bounces creates a local slope under the droplet which compensate for the centrifugal inertia. Using a continuous approximation model, they found that spinning could be maintained even for a vanishing Coriolis force but for parameters non accessible to experimental conditions [25,26]. The same conclusion was drawn by a stability analysis performed with a discrete-time model based on first principles in a specific parameter regime [28,29]. Spin states were observed with walkers released from a harmonic potential but only for limited time [30]. In all these cases, the spinning however did not occur spontaneously. Here, we explore a new regime of small and slow droplets for which stable spin states emerge spontaneously.

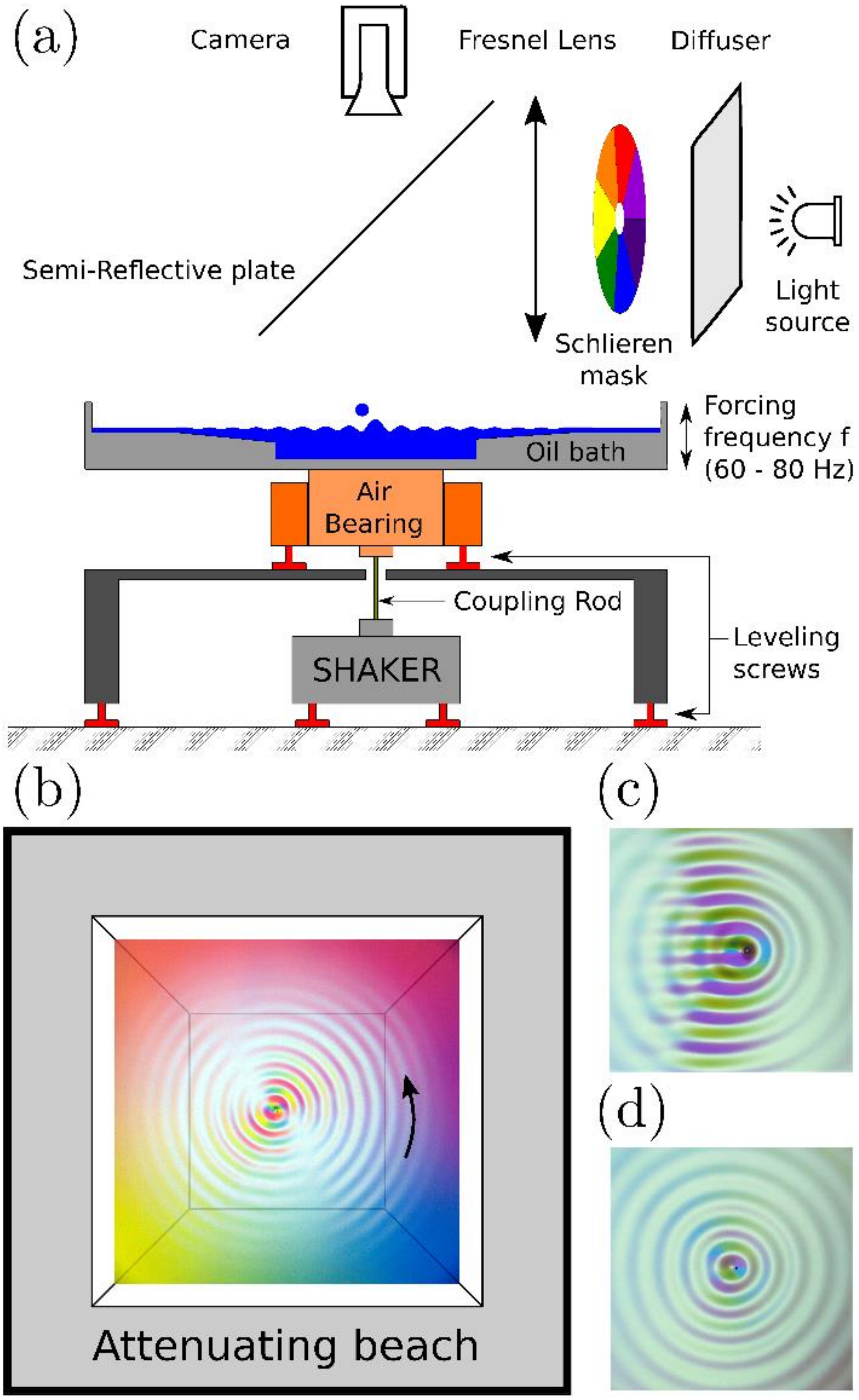

***Fig. 1*** *(a) Schematics of the experimental setup. A color Schlieren technique encodes the slope direction of the wave field. (b) Typical snapshot of the wave field and its associated droplet superimposed with a top view graphic of the container. Walkers going in a straight line (c) large fast walker (mean speed $v = 10$ mm.s$^{-1}$ and $D_w = 0.57$ mm) and (d) small slow walker (mean speed $v = 2$ mm.s$^{-1}$ and $D_w = 0.43$ mm), $Me = 40$ and $\nu_{exc} = 72$ Hz*

## 2 Experimental setup

Figure 1a shows a schematic of the experimental setup. The container is fixed on a shaker (LDS V455) vibrating vertically at frequency $\nu_{\mathrm{exc}}$ between 60 Hz and 80 Hz. A linear air bearing is used to avoid transverse vibrations [31]. The whole experiment is confined in a Plexiglass box to limit air flows. The 3D printed square container consists of a 60 mm wide, 8.5 mm deep square cavity surrounded by a 55 mm wide shallow bank. This profile is optimized by try and error method to attenuate the waves emitted by the walls using a 0.5 mm deep shallow liquid section and minimize wave reflection from the cavity using a 30 mm wide, 2 mm high chamfer. The amplitude of the reflected wave for typical experimental parameters is smaller one micron. The cavity is filled with silicon oil of viscosity $\mu = 20 \cdot 10^{-3}$ Pa.s, density $\rho = 950$ kg.m$^{-3}$ and surface tension $\gamma = 0.0206$ N.m$^{-1}$. The droplets are produced by a piezoelectric generator similar to the one in ref. [32]. The droplets have a tunable diameter from 0.4 mm to 1 mm with a 10 µm precision.

The illumination is performed with a radial colored Schlieren mask projected on the liquid surface using a Fresnel lens which enables color-coding the direction of the surface slope [33]. The camera placed above the container is synchronized with the Faraday frequency $\nu_{\mathrm{F}} = 1/T_{\mathrm{F}} = \nu_{\mathrm{exc}}/2$ with a tunable relative phase to tune the recorded amplitude of the wave field. The bouncing droplets which have undergone period doubling are strobed at the bouncing frequency. Their trajectories are retrieved from image analysis. Figure 1b shows a typical snapshot with the bouncing droplet with its wave field superimposed on a schematic top view of the cavity. The wave field surrounding the droplet extends over 10 Faraday wavelengths $\lambda_{\mathrm{F}}$, well beyond the deep region of the cavity with no detectable field reflection nor field discontinuity in the shallow region. This ensures that the influence of the size and boundaries of the cavity on the walker dynamics is limited. Each bouncing of the droplet on the shaken liquid surface excites locally the Faraday instability which is maintained during a characteristic time $\tau_{\mathrm{Me}}$. It is convenient to normalized $\tau_{\mathrm{Me}}$ by the Faraday period $T_{\mathrm{F}}$ to define a memory parameter $Me = \tau_{\mathrm{Me}}/T_{\mathrm{F}}$ for a walker which characterizes the number of previous bounces that contribute to the global wave field. The latter can be tuned using its dependence to the Faraday threshold. $\tau_{Me}$ satisfies $Me = (1 - \gamma/\gamma_F)^{-1}$ with $\gamma$ and $\gamma_F$ being the forcing acceleration and the Faraday acceleration threshold respectively. The measurement of $\tau_{Me}$ can be performed simply by inducing a local perturbation on the liquid surface and measuring the exponential decay of the excited Faraday wave field. $\tau_{Me}$ is measured experimentally by inducing a point perturbation on the liquid surface and fitting the exponential decay of the excited Faraday waves. This gives $Me$ with an uncertainty of typically 10% (in the range of the study).

Small and slow droplets should be good candidates for spin states since the centrifugal force is proportional to their mass and to the square of their speed. The minimal size is set by the vertical dynamics of the droplet [34–36]. While most studies use a typical droplet diameter of $D_{\mathrm{w}} = 0.7$ mm, we experimentally found the period doubled bouncing regime could be maintained down to droplet diameter of $D_{\mathrm{w}} = 0.43$ mm for $\nu_{\mathrm{exc}} = 70$ Hz. Small droplets are much slower than large ones [34], typically $v = 2$ mm.s$^{-1}$ versus $v = 10$ mm.s$^{-1}$. Figure 1c and 1d shows the associated wave field for a straight trajectory with $\lambda_{\mathrm{F}} = 5.2$ mm at $Me = 40$. Fast droplets create typical interference patterns resulting from a memory length $l_{\mathrm{Me}} = v\,\tau_{\mathrm{Me}}$ persistent over several $\lambda_F$ [16,18,37]. For slow droplets, $l_{\mathrm{Me}} \approx \lambda_{\mathrm{F}}$ resulting in a nearly circular wave field. Although their characteristics increase the possibility of an equilibrium, these slow walkers have not been investigated in previous studies on spin state.

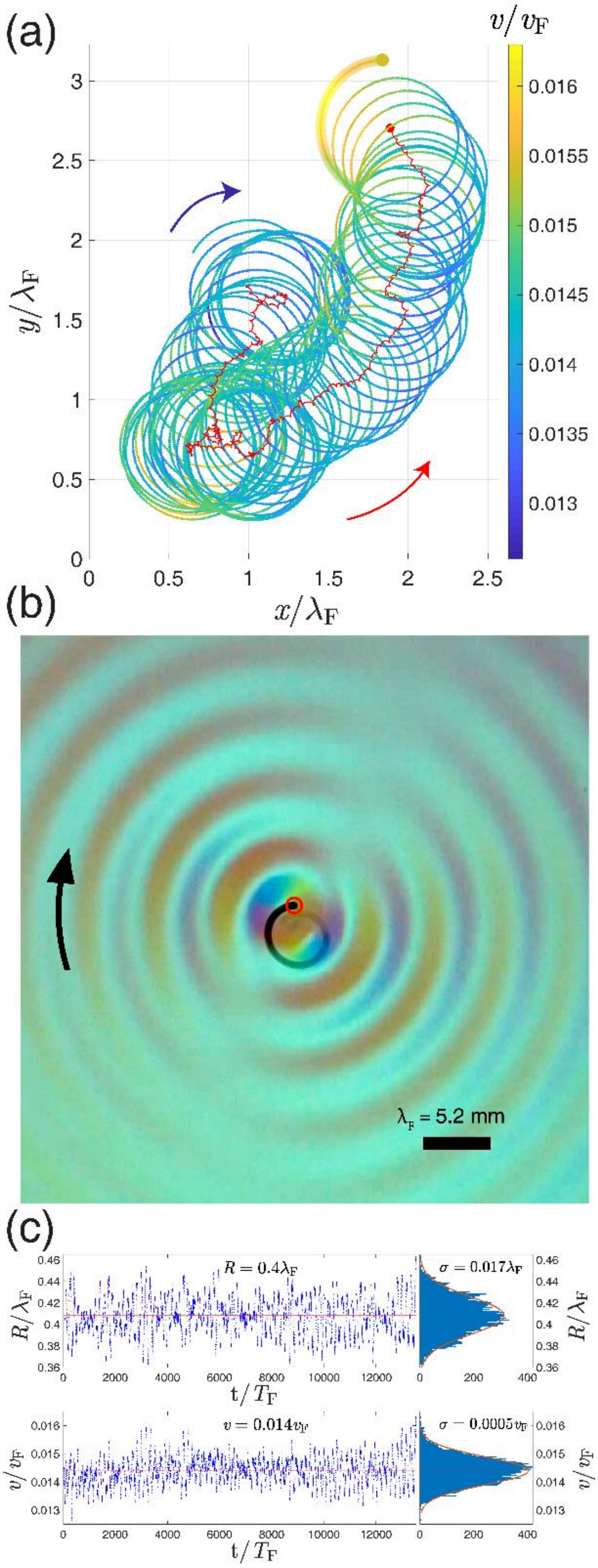

***Fig. 2*** *(a) Trajectory of a walker in a clockwise spin state with color coded speed ($Me = 70$, $\nu_{exc} = 70$ Hz, $D_w = 0.46$ mm). The bold final segment represents the memory length $l_{Me} =$*

*$v\,\tau_{Me}$. Trajectory of the center of curvature (red line). (b) Snapshot of the associated wave field and the superimposed trajectory (black line) with transparency proportional to attenuation (see Online Movie 1). (c) Time evolution of the instantaneous orbital radius $R/\lambda_F$ and speed $v/v_F$ and their associated distributions with standard deviation $\sigma_R$ and $\sigma_v$ respectively.*

**Experimental results**

Figure 2a shows a typical spin state trajectory with 76 revolutions for a droplet with $D_{\mathrm{w}} = 0.46$ mm and $Me = 70$ (see Online Movie 1). The center of the trajectory (red line) is obtained by fitting the last *Me* bounces positions along the trajectory with a circle. Typically, droplets enter spinning trajectories either spontaneously or after a small perturbation. Undetectable noise fluctuations seem to be sufficient to trigger the spin state in contrast with previous studies for which the droplet needed to be set into the spin state [30]. Self-spinning can occur anywhere in the cavity before slowly drifting in a random direction in agreement with the negligible influence of the boundaries. The spontaneous breaking of symmetry into stable spin states must ultimately be induced by (undetected) very small experimental noise to be reached. Although air currents are minimized by the use of plexiglass walls, it is most probable that they play a role in the drift (especially due to the very small size of the droplets). For this particular set of parameters, the orbiting period is approximately twice the memory time $\tau_{\mathrm{Me}}$ ($l_{\mathrm{Me}}$ is in bold line in Fig. 2a). Stable spin state trajectories of several hundred rotations are observed within the approximate narrow range of parameters of forcing frequencies $70\ \mathrm{Hz} \leq \nu_{\mathrm{exc}} \leq 76$ Hz, wave field memories $70 \leq Me \leq 100$ and droplet diameters $0.43\ \mathrm{mm} \leq D_{\mathrm{w}} \leq 0.56$ mm. The characteristics of the observed spin states are similar within this parameter range with an increase destabilization of the spin states as the droplet size or the speed of the droplet increase. As the droplet size increases, the average time spent in the spin states decreases and random exit events appears after which the droplet wanders in the cavity before returning to a spin state (see Online Movie 2). The small range of parameter is defined by the observation of spin states orbiting for more than 10 $\tau_{Me}$. The limit at small droplet size is set by the vertical dynamics of the droplet which undergoes a transition towards a chaotic bouncing regime below $D_{\mathrm{w}} \leq 0.43$ mm [34–36].

Figure 2b is a typical snapshot of a walker in a spin state. The transparency of the trajectory line indicates the vanishing contribution of successive impacts to the wave field. This wave field has a spatio-temporal helical symmetry and can be decomposed primarily into a zero and a first order Bessel functions centered on the orbit [30]. The former creates a local axial slope $\nabla h^{\perp}(\boldsymbol{r})$ at the droplet position $\boldsymbol{r}$, which gives a centripetal force balancing the centrifugal inertial effects. The latter rotates with the droplet creating a local tangential slope $\nabla h^{\parallel}(\boldsymbol{r})$ for self-propulsion. Figure 2c shows the time evolution of the instantaneous orbital radius $R/\lambda_{\mathrm{F}}$ and the speed $v/v_{\mathrm{F}}$ of the walker normalized by the Faraday wavelength $\lambda_{\mathrm{F}}$ and the wave phase velocity $v_{\mathrm{F}} = \lambda_{\mathrm{F}}/T_{\mathrm{F}}$ respectively. Both exhibit small oscillations at the orbital period with mean values $R = 0.4\lambda_{\mathrm{F}}$ and $v = 0.014v_{\mathrm{F}}$ respectively. Their distributions have similar Gaussian profiles with a narrow standard deviation ($\sigma_R/R = 4.3\%$ and $\sigma_v/v = 3.6\%$ respectively). Their fluctuations are correlated (with a linear correlation coefficient of 0.75) indicating a coupling between the droplet propulsion and the radial force provided by the wave field.

## 3 Model of the spinning dynamics

The walker dynamics are described by a simple discrete path-memory model which takes the essence of the dynamics [18,19,30]. The vertical and horizontal motion of the walkers are decoupled. Walkers undergo parabolic jumps between inelastic bounces. Following Durey *et al.* [28,29], we consider the bouncing time as instantaneous and the horizontal velocity $\boldsymbol{v}_{n+1}$ after the $n^{\mathrm{th}}$ bounce constant. The position of the $(n+1)^{\mathrm{th}}$ bounce $\boldsymbol{r}_{n+1}$ is given by iteration: $\boldsymbol{r}_{n+1} = \boldsymbol{r}_n + \boldsymbol{v}_{n+1}T_F$. The impact induces a change in the horizontal velocity satisfying

$$\frac{m}{T_F}(\boldsymbol{v}_{n+1} - \boldsymbol{v}_n) = -D\frac{(\boldsymbol{v}_{n+1}+\boldsymbol{v}_n)}{2} - C\boldsymbol{\nabla}h(\boldsymbol{r}_n) \quad (1)$$

with $m$ being the droplet mass, $\boldsymbol{\nabla}h(\boldsymbol{r}_n)$ the surface gradient at $\boldsymbol{r}_n$, $C$ and $D$ the wave coupling and the friction constant respectively. The change in velocity is driven by a friction-like force originating mainly in the shearing of the air layer between the drop and the bath, and a wave or memory force, proportional to the local surface slope. Neglecting viscosity [18] and wave propagation, the wave field $h(\boldsymbol{r})$ can be described as a superposition of $J_0$ Bessel functions with initial amplitude $h_0$ centered at the impact positions along the trajectory and decaying exponentially with a memory time $\tau_{\mathrm{F}}$. At the time of the $n^{\mathrm{th}}$ impact, it satisfies

$$h(\boldsymbol{r}) = h_0 \sum_{q=1}^{\infty} J_0\left(k_{\mathrm{F}}\left|\boldsymbol{r} - \boldsymbol{r}_{n-q}\right|\right)\exp\left(-\frac{q}{Me}\right) \quad (2)$$

The dynamics of the walkers are thus fully characterized by the two parameters $Ch_0$ and $D$. For a walker of mass $m$ and speed $v$ in an orbit of radius $R$, eq. (1) projected along the radial and tangential directions gives respectively

$$F_{\perp}(R,v) = \frac{mv^2}{R} - C\nabla h^{\perp}(\boldsymbol{r}_n) \quad (3)$$

$$F_{\parallel}(R,v) = -Dv - C\nabla h^{\parallel}(\boldsymbol{r}_n) \quad (4)$$

For the spin state values $(R, v)$, the force balance writes $F_{\perp}(R,v) = 0$ and $F_{\parallel}(R,v) = 0$. The centripetal wave force balances the inertial force involving the parameter $Ch_0$ only (eq. 3). It equals to $1.56 \cdot 10^{-10}$ N in the present case (fig. 2, with $Me = 70$, $\nu_{\text{exc}} = 70$ Hz, $D_{\text{w}} = 0.46$ mm), a value approximately 25 times smaller than that of "standard" walkers. The tangential component yields the parameter $D$ (eq. 4).

Considering the walker in a given circular trajectory, the bouncing positions are on the vertices of a regular polygon from which $\nabla h^{\perp}(\boldsymbol{r}_n)$ and $\nabla h^{\parallel}(\boldsymbol{r}_n)$ can be computed. From the experimental observations, we obtain $Ch_0 = 2.22 \cdot 10^{-13}$ N.m and $D = 1.03 \cdot 10^{-6}$ kg.s$^{-1}$ which are values in the range of those calculated from the hydrodynamic model [26].

Decoupling the parameters in eq (3) and (4) allows experimental calibration using measurements on spin states. All the proposed models ultimately determine two parameters associated with the wave propulsion and the damping force. However, the diversity of hypotheses and formalisms make it difficult to find an exact correspondence between these parameters. An experimental calibration would make comparison possible. In addition, when these parameters are rooted in the hydrodynamic of the bouncing, this calibration would provide an additional validation of the hydrodynamic hypothesis.

We use the calibrated values to run simulations based on eq. (1) and (2) and observe the outcome on the walker dynamics with the first $N$ impacts placed on a spin state as initial conditions. For $Me = 70$ and $N > 75$, the walker enters a stable spin state in agreement with the experimental findings and validating the calibrations *a posteriori*.

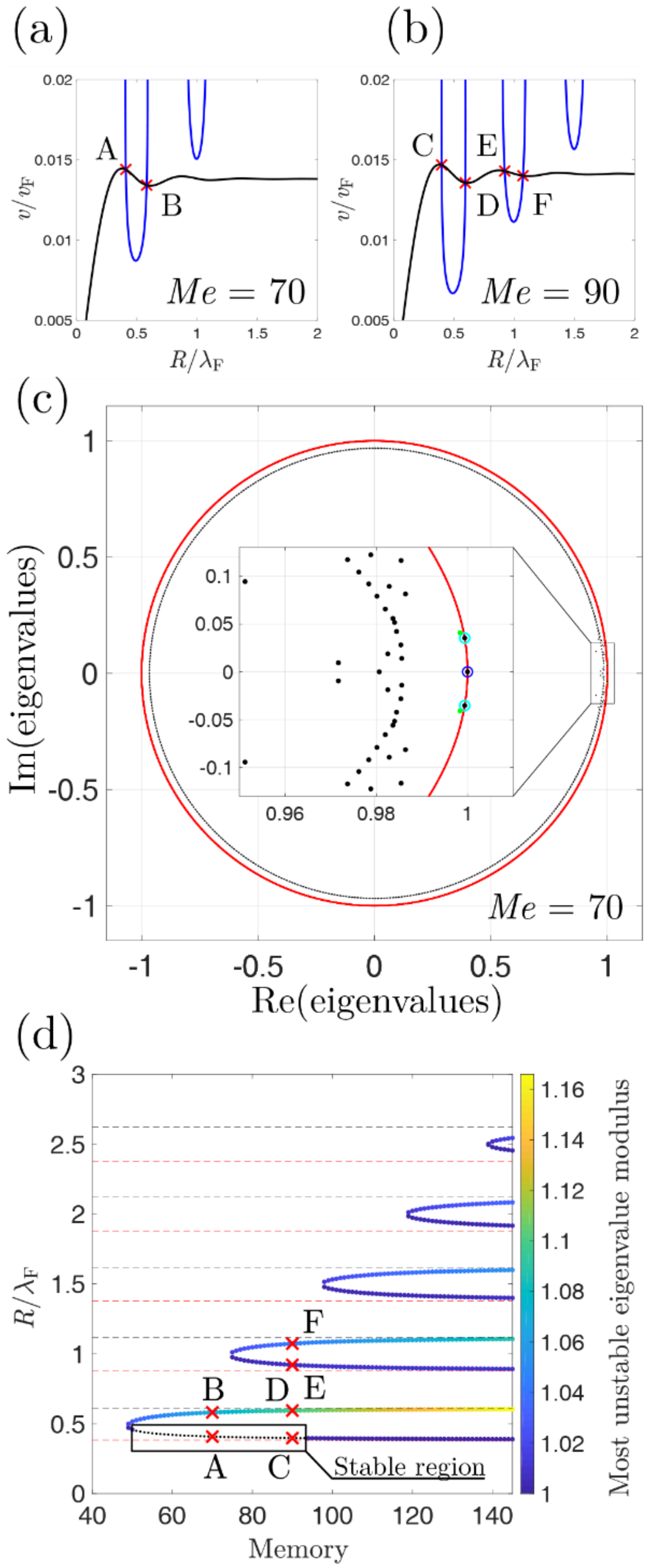

***Fig. 3*** *Radial force equilibrium* $F_\perp(R, v) = 0$ *(blue line) and tangential one* $F_\parallel(R, v) = 0$ *(black line) in the* $(R, v)$ *plane for* $Me = 70$ *(a) and* $Me = 90$ *(b). Intersection points (crosses)*

*associated to possible spin states labeled A, B and C, D, E, F respectively. (c) Eigenvalues (black dots) for the discrete linear stability matrix around equilibrium point A (inset: close up). Two eigenvalues associated with the translational invariance (blue circles), the rotational one (cyan circle) and the least-stable eigenvalue (green points). (d) Modulus of the most-unstable eigenvalue (color coded) of the spin state in the* $(R/\lambda_F, Me)$ *plane.* $J_0(k_F R) = 0$ *(dashed red lines) and* $J_1(k_F R) = 0$ *(dashed black lines). Using values from the calibration, we explore the existence and stability of the spin states. These modes characterized by* $(R, v)$ *must simultaneously satisfy* $F_\perp(R, v) = 0$ *and* $F_\parallel(R, v) = 0$*. Figure 3a and 3b show the two curves (blue and black lines respectively) in the* $(R, v)$ *plane for* $Me = 70$ *and* $Me = 90$ *respectively. The radial wave force is mainly associated to a centered Bessel mode* $J_0(k_F r)$ *with an excitation amplitude proportional to* $MeJ_0(k_F R)$ *at high memory[30]. Hence, the radial wave force exerted on an orbit of radius R is proportional to* $MeJ_0(k_F R)J_1(k_F R)$*. As* $Me$ *increases, the condition* $F_\perp(R, v) = 0$ *is asymptotically fulfilled by* $J_0(k_F R) \simeq 0$ *or* $J_1(k_F R) \simeq 0$*. The tangential wave force is mainly associated with a centered* $J_1$ *Bessel function rotating with the droplet. For* $R \gtrsim 0.25\lambda_F$*, the driving force balances the shearing force resulting in a nearly constant droplet speed with a mean value of* $v/v_F = 0.014$*. Two possible spin states can be found for* $Me = 70$ *(labeled A and B in Fig. 3a). Couples of new possible modes appear as* $Me$ *increases, e.g. four modes appear for* $Me = 90$ *(see Fig. 3b).*

## 4 Discrete mode stability analysis

We use the discrete iterative path-memory model to assess the stability of the modes [28,29]. We reformulate the walker dynamics as a non-linear iterative map on a $p$ dimensional complex space. The walker dynamics are fully determined at the time of the $n^{\text{th}}$ impact by the previous ones. The walker state is given by a vector $\boldsymbol{Z}_n$ containing the positions of the previous impacts written in the complex plane. The $q^{\text{th}}$ component of the vector $\boldsymbol{z}_n^q$ contains the $(q-1)^{\text{th}}$ impact, $\boldsymbol{z}_n^q = \boldsymbol{r}_{n-q+1} \in \mathbb{C}$ with $q \in \mathbb{N}^*$, $\boldsymbol{z}_n^1$ being the position of the $n^{\text{th}}$ impact. The walker dynamics is calculated iteratively using a function $f$ such that $\boldsymbol{Z}_{n+1} = f(\boldsymbol{Z}_n)$ (see Online Supplementary materials 2). In practice, the exponential wave field decreases with a characteristic time $\tau_{Me}$, $\boldsymbol{Z}_n$ can be truncated to the $p$ previous impacts satisfying $e^{-p/Me} \sim 0.1\%$.

A walker with a state vector $\boldsymbol{Z}_n$ in a spin state $(R, v)$ is rotated at each step. Hence, $f(\boldsymbol{Z}_n) = rot_\theta(\boldsymbol{Z}_n)$, $rot_\theta$ being a rotation in $\mathbb{C}^p$ by an angle $\theta = 2\,\text{asin}(vT_\text{F}/2R) \approx vT_\text{F}/R$. $\boldsymbol{Z}_n$ is thus a fixed point of the nonlinear operator $g = rot_{-\theta} \circ f$. The linear stability analysis of the spin state can be performed using $g$ at point $\boldsymbol{Z}_n$. The mode is stable if the eigenvalues of the Jacobian matrix of $g$ evaluated at $\boldsymbol{Z}_n$ are within the unit circle. The eigenvalues outside are linearly unstable if real and oscillatory unstable otherwise [38]. Figure 3c shows the eigenvalues found for $Me = 70$ associated with point A (Fig. 3a) with $Ch_0$ and $D$ calibrated experimentally. These values are kept constant over the small varying range of $Me$ since no appreciable changes can be measured on the experimental characteristics of the spin states. This agrees with the hydrodynamic-based models [25,39].

The mode is stable with all the eigenvalues inside the unit circle. The eigenvalues associated with translational (blue circles) and rotational invariance (cyan circle) are discarded for the stability analysis. Figure 3d shows the stability map of the possible spin states of radius $R/\lambda_F$ versus $Me$ given by the highest eigenvalue norm. Modes labeled in Fig. 3a and 3b are plotted. The orbits associated with the centered $J_0$ Bessel function are stable for $Me$ spanning from 51 to 93 (modes $A$ and $C$). All the other modes are found unstable, meaning that only the smallest orbits are stable in agreement with the experimental findings (see fig. 2). The instability associated with the upper (resp. lower) branches of the curves is linear (resp. oscillatory). At high memory ($Me >$100), the orbits appear experimentally unstable. It is interesting to note that in the case of spin states obtained in rotating frames a similar destabilization occurs at high memory experimentally [27] and theoretically [39]. The limited cavity size might also play a role in perturbing the walker for such extended wave fields.

## 5 Discussion

A general feature of walker's dynamics is that at intermediate memory values (typically for $Me$ equals a few tens) the trajectories tends to self-organize with their wave field into pivotal structures around which the walkers acquire curved trajectories with preferred curvature such as in the spin states [12]. The straight trajectories of the walkers must however be initially perturbed to enter these curved paths since straight trajectories are stable [17]. The reflections of the walker on the cavity boundaries breaks the translational symmetry and enables the walker revisit the wave field it as produced during the previous bounces. However, in the case of "standard" walkers, this wave field is not sufficient to create pivotal structures which would induce more complex behavior. The walker recovering a straight trajectory after its interaction with the wall [16]. More complex trajectories with the emergence of preferred curvatures are only observed when the interaction with the wave field is more important obtained by increasing the memory, the confinement of the walker to small region [13,40] or the reflections of a cavity [41]. In the present case for slow and small droplets, the disturbance caused by the wave field on the walker's trajectory is more significant and the droplet undergoes meandering trajectories even in the presence of a relatively small wave field. In addition, these walkers with curved trajectories can spontaneously enter self-organize into a stable spin state. In the case of "standard" walker this is possible only if the droplets are initially prepared into the spin state [26,30]. This results from the fact that the wave field needs to organized in a more accurate way to compensate for the stronger centrifugal force. Even then, the stability of these spin states is still much more tenuous compared to small and slow droplets. This is the reason why such droplets can “spontaneously” enter in a stable spin state anywhere under the influence of a perturbating wavefield. This wavefield can automatically be produced by the reflection of the walker on the cavity boundaries which forces the droplet to revisit its own wave field.

It is interesting to compare the present stability results based on a discrete model with the ones based on the continuous approximation using integro-differential equations [25,26]. Using the experimental calibration, the spin state is found unstable ($\kappa_0 = 0.29$) but very close to stability ($\kappa_0 \lesssim 0.2$, with $C = 0.33$, $\sin\Phi = 0.16$). Small droplets could slightly change the hydrodynamics parameters to attain stability.

The discrete stability analysis can also be applied to “standard” walkers in previous experiments for which orbits could be maintained only for a limited times [30]. The parameter values are retrieved from the calibration method ($Ch_0 = 8.8 \cdot 10^{-12}$ N. m and $D = 7.2 \cdot 10^{-6}$ kg. s$^{-1}$). The discrete stability analysis finds stability for a range of intermediate memories. For high memories, simulations show that the spin state is wobbling but maintained while it destabilizes at even higher memories. The model of Durey et al. [28] based on first principles with an instantaneous contact approximation could possibly find spin state stability with this new set of parameters. The finding of the new stable spin state for walkers together with the proposed calibration method should definitely help compare and rationalize the many walking models with their quite diverse hypotheses from continuous to instantaneous approximations.

**Acknowledgments:** In memory of Yves Couder who would have enjoyed observing these spin states. The authors are grateful to Antonin Eddi, Sander Wildeman and Chloé d’Hardemare for insightful discussions. The authors thank the support of AXA research fund the French National Research Agency LABEX WIFI (ANR-10-LABX-24), and a Freeside Fund #CoS grant.

The datasets generated during and/or analyzed during the current study are available from the corresponding author on reasonable request.

**Supplementary Material**

# Spontaneous emergence of a spin state for an emitter in a time-varying medium.

Samuel Bernard-Bernardet,* Marc Fleury,† and Emmanuel Fort*

A discrete path-memory model for the walking droplet is detailed, where its dynamics is reformulated as a non-linear iterative map on a p dimensional complex space. The linear stability analysis of the self-spinning mode is performed by computing the eigenvalues of the jacobian matrix of the suitable operator.

## I. NON-LINEAR DISCRETE P-DIMENSIONAL MODEL

For a system with Memory Me, only a certain number of impacts along the droplet trajectory have a significant influence on the field. Indeed, the local Faraday mode excited at each impact decays exponentially and its influence can be considered to vanish after a certain time. We consider that the wave field created by a single impact is vanishing if its amplitude is less than 0.1 percent of the maximum amplitude.

The walking droplet motion is modeled by the dynamics of the discrete set of the last p impacts of the droplets. p increases with the memory of the system.

The principle of this model is described in [1]

At iteration n, the positions of the previous $p^{th}$ impacts are :

$$\vec{z}_n^p = \begin{bmatrix} x_n^p \\ y_n^p \end{bmatrix} \tag{1}$$

With $\vec{z}_n^1$ the position of the droplet at iteration n.

For clarity, we will consider $\vec{z}_n^p = x_n^p + iy_n^p \in \mathbb{C}$ and explicit the dynamics as the evolution of a finite set of points in the complex plane.

$\vec{z}_n^1$ is the position of the droplet just before the impact at iteration n, and its speed just before it gets an impulse from the field is :

$$\vec{v}_n^1 = \frac{\vec{z}_n^1 - \vec{z}_n^2}{T_F} \tag{2}$$

Where $T_F$ is the Faraday period. The position of the next impact is :

$$\vec{z}_{n+1}^1 = \vec{z}_n^1 + T_F \vec{v}_{n+1}^1 \tag{3}$$

where the speed $\vec{v}_{n+1}^1$ just after the current $n^{th}$ impact is calculated following a discrete impulse from the field :

$$m\frac{(\vec{v}_{n+1}^1 - \vec{v}_n^1)}{T_F} = -D(\frac{\vec{v}_{n+1}^1 + \vec{v}_n^1}{2}) - C\vec{\nabla}h(\vec{z}_n^1) \tag{4}$$

The discrete impulse at impact n has 2 components.

- The viscous dissipation during contact time $-D(\frac{\vec{v}_{n+1}^1 + \vec{v}_n^1}{2})$ is equally divided before and after the field impulse.

- The field contribution $-C\vec{\nabla}h(\vec{z}_n^1)$ is its slope at position of impact n, computed from the previous impacts, according to :

$$\vec{\nabla}h(\vec{z}_n^1) = -h_0 k_F \sum_{k=2}^{p} \frac{J_1(k_F\|\vec{z}_n^1 - \vec{z}_n^k\|)}{\|\vec{z}_n^1 - \vec{z}_n^k\|}(\vec{z}_n^1 - \vec{z}_n^k)e^{-(k-1)/Me} \tag{5}$$

* Institut Langevin, ESPCI Paris, Université PSL, CNRS, 1 rue Jussieu, 75005, Paris, France

† Two Prime, Open Finance Group, Atlanta, USA

where $h_0$ is the maximum amplitude of the wave field generated by one impact.
This yields the following recurrence formula for the position of the droplet :

$$\vec{z}_{n+1}^{1} = I\vec{z}_n^1 + J\vec{z}_n^2 + K\sum_{k=2}^{p}\frac{J_1(k_F\|\vec{z}_n^1 - \vec{z}_n^k\|)}{\|\vec{z}_n^1 - \vec{z}_n^k\|}(\vec{z}_n^1 - \vec{z}_n^k)e^{-(k-1)/Me} \tag{6}$$

$$\text{with } I = 1 + \frac{(1-\frac{DT_F}{2m})}{(1+\frac{DT_F}{2m})}, J = 1 - I, \text{ and } K = \frac{Ch_0T_F^2k_F}{m(1-\frac{DT_F}{2m})}$$

The position of all other previous impacts is updated according to a shift :

$$\vec{z}_{n+1}^{j} = \vec{z}_n^{j-1} \text{ for } j >= 2 \tag{7}$$

We now write the state vector as a p dimensional vector in $\mathbb{C}^p$:

$$\vec{Z}_n = \begin{bmatrix} \vec{z}_n^1 \\ \vdots \\ \vec{z}_n^j \\ \vdots \\ \vec{z}_n^p \end{bmatrix} \tag{8}$$

The evolution of the state vector ids given by f : $\mathbb{C}^p \rightarrow \mathbb{C}^p$

$$\vec{Z}_{n+1} = f(\vec{Z}_n) = f(\vec{z}_n^1, \ldots, \vec{z}_n^p) \tag{9}$$

that is :

$$\begin{bmatrix} \vec{z}_{n+1}^1 \\ \vdots \\ \vec{z}_{n+1}^j \\ \vdots \\ \vec{z}_{n+1}^p \end{bmatrix} = f(\begin{bmatrix} \vec{z}_n^1 \\ \vdots \\ \vec{z}_n^j \\ \vdots \\ \vec{z}_n^p \end{bmatrix}) = \begin{bmatrix} f_1(\vec{z}_n^1, \ldots, \vec{z}_n^p) \\ \vdots \\ f_j(\vec{z}_n^1, \ldots, \vec{z}_n^p) \\ \vdots \\ f_p(\vec{z}_n^1, \ldots, \vec{z}_n^p) \end{bmatrix} \tag{10}$$

From equations (6) we get :

$$f_1(\vec{z}_n^1 \ldots, \vec{z}_n^p) = I\vec{z}_n^1 + J\vec{z}_n^2 + K\sum_{k=2}^{p}\frac{J_1(k_F\|(\vec{z}_n^1 - \vec{z}_n^k))\|)}{\|(\vec{z}_n^1 - \vec{z}_n^k))\|}(\vec{z}_n^1 - \vec{z}_n^k)e^{-(k-1)/Me} \tag{11}$$

For $2 \leq j \leq p$, we get from (11) :

$$f_j(\vec{z}_n^1, \ldots, \vec{z}_n^p) = \vec{z}_n^{j-1} \tag{12}$$

This completes the description of the discrete dynamical system as a non-linear operator f acting on $\mathbb{C}^p$.

## II. CIRCULAR ORBIT AND ROTATION OPERATOR

Let $(\vec{s_n})$ be a circular solution $(\vec{s_{n+1}} = f(\vec{s_n}))$ associated with radius $R$ and pulsation $\Omega$. At each iteration, each point of the state vector rotates by a constant angle $\theta = \Omega T_F$ :

$$\vec{s_n} = \begin{bmatrix} Re^{(in\Omega T_F)} \\ \vdots \\ Re^{(i(n-p+1)\Omega T_F)} \end{bmatrix} = \begin{bmatrix} Re^{in\theta} \\ \vdots \\ Re^{i(n-p+1)\theta} \end{bmatrix} \tag{13}$$

We now define a rotation operator $rot_\theta$ acting on the state vector, which rotates each component $\vec{z}_n^j (j = 1 \dots p)$ by an angle $\theta$ :

$$rot_\theta(\begin{bmatrix} \vec{z}_n^1 \\ \vdots \\ \vec{z}_n^p \end{bmatrix}) = \begin{bmatrix} e^{i\theta}\vec{z}_n^1 \\ \vdots \\ e^{i\theta}\vec{z}_n^p \end{bmatrix} \tag{14}$$

That is : $rot_\theta(\vec{Z}_n) = e^{i\theta}\vec{Z}_n$

We notice that $\vec{s_{n+1}} = rot_\theta(\vec{s_n})$, hence $\vec{s_n}$ is a fixed point of the non linear operator g defined by $g = rot_{-\theta} \circ f$ for any n.

We also note from (11) and (12) that :

$$rot_{-\theta} \circ f = f \circ rot_{-\theta} \tag{15}$$

and since f and $rot_{-\theta}$ commute, so do g and $rot_\theta$.

This is the so called *equivariance* condition [2]. This commutativity condition will allow us to deduce the stability criteria of the circular orbits from the stability analysis of the fixed point $\vec{s}$ of operator g.

## III. STABILITY ANALYSIS

We now consider the state vector as part of $\mathbb{R}^{2p}$ instead of $\mathbb{C}^p$, and modify the operators f, g and $rot_\theta$ accordingly. This will allow us to compute Jacobian matrices.

Let's call $\vec{s}$ a fixed point of operator g, corresponding to a state vector taken at any iteration of a circular solution of dynamical system f. For a small perturbation $\epsilon$ around s, we have at first order :

$$g(\vec{s} + \epsilon) \approx g(\vec{s}) + J_g(\vec{s}).\epsilon = \vec{s} + J_g(\vec{s}).\epsilon$$

where $J_g(\vec{s})$ is the Jacobian of operator g at $\vec{s}$. Hence :

$$g^{(n)}(\vec{s} + \epsilon) \approx \vec{s} + J_g^n(\vec{s}).\epsilon \tag{16}$$

If the eigenvalues of $J_g(\vec{s})$ all have a modulus strictly inferior to 1, s is a stable stationary point for g. Now, let's look at the relative evolution of a slightly perturbed circular solution under $f$

$$\begin{aligned}
&f^{(n)}(\vec{s} + \epsilon) - f^{(n)}(\vec{s}) \\
&= (rot_\theta \circ g(\vec{s} + \epsilon))^{(n)} - (rot_\theta \circ g(\vec{s}))^{(n)} \\
&= rot_\theta^{(n)} \circ g^{(n)}(\vec{s} + \epsilon) - rot_\theta^{(n)} \circ g^{(n)}(\vec{s}) \text{ (because } g \text{ and } rot_\theta \text{ commute)} \\
&= rot_{n\theta} \circ (\vec{s} + J_g^n(\vec{s}).\epsilon) - rot_{n\theta}(\vec{s}) \text{ (from (16))} \\
&= rot_{n\theta}(J_g^n(\vec{s}).\epsilon)
\end{aligned} \tag{17}$$

The norm of the difference between a state vector corresponding to the stable circular trajectory and one corresponding to the perturbed trajectory evolves according to $J_g(\vec{s})$.

In other words, the stability of a circular trajectory under f is given by the eigenvalues of the Jacobian of $g = rot_{-\theta} \circ f$ taken at any state vector in the stable circular trajectory.

$$J_g(\vec{s}) = J_{rot_{-\theta} \circ f}(\vec{s}) = J_{rot_{-\theta}}(f(\vec{s}))J_f(\vec{s}) = rot_{-\theta}.J_f(\vec{s}) \tag{18}$$

To compute the Jacobian $J_f(\vec{s})$, we now need to write explicitly the state vector as a 2p dimensional vector :

$$\vec{r}_n = \begin{bmatrix} x_n^1 \\ y_n^1 \\ \vdots \\ x_n^i \\ y_n^1 \\ \vdots \\ x_n^p \\ y_n^p \end{bmatrix} \tag{19}$$

The dynamics is now an evolution given by f : $\mathbb{R}^{2p} \to \mathbb{R}^{2p}$

$$\vec{r}_{n+1} = f(\vec{r}_n) = f(x_n^1, y_n^1, \ldots, x_n^i, y_n^i \ldots, x_n^p, y_n^p) \tag{20}$$

that is :

$$\begin{bmatrix} x_{n+1}^1 \\ y_{n+1}^1 \\ \vdots \\ x_{n+1}^i \\ y_{n+1}^i \\ \vdots \\ x_{n+1}^p \\ y_{n+1}^p \end{bmatrix} = f(\begin{bmatrix} x_n^1 \\ y_n^1 \\ \vdots \\ x_n^i \\ y_n^i \\ \vdots \\ x_n^p \\ y_n^p \end{bmatrix}) = \begin{bmatrix} f_1(x_n^1, y_n^1, \ldots, x_n^j, y_n^j \ldots x_n^p, y_n^p) \\ f_2(x_n^1, y_n^1, \ldots, x_n^j, y_n^j \ldots x_n^p, y_n^p) \\ \vdots \\ f_{2i-1}(x_n^1, y_n^1, \ldots, x_n^j, y_n^j \ldots x_n^p, y_n^p) \\ f_{2i}(x_n^1, y_n^1, \ldots, x_n^j, y_n^j \ldots x_n^p, y_n^p) \\ \vdots \\ f_{2p-1}(x_n^1, y_n^1, \ldots, x_n^j, y_n^j \ldots x_n^p, y_n^p) \\ f_{2p}(x_n^1, y_n^1, \ldots, x_n^j, y_n^j \ldots x_n^p, y_n^p) \end{bmatrix} \tag{21}$$

From equations (6) we get :

$$\begin{aligned} & f_1(x_n^1, y_n^1, \ldots, x_n^j, y_n^j \ldots, x_n^p, y_n^p) \\ & = I x_n^1 + J x_n^2 + K \sum_{k=2}^{p} \frac{J_1(k_F \sqrt{(x_n^1 - x_n^k)^2) + (y_n^1 - y_n^k)^2)}}{\sqrt{(x_n^1 - x_n^k)^2) + (y_n^1 - y_n^k)^2}} (x_n^1 - x_n^k) e^{-(k-1)/Me} \end{aligned} \tag{22}$$

$$\begin{aligned} & f_2(x_n^1, y_n^1, \ldots, x_n^j, y_n^j \ldots, x_n^p, y_n^p) \\ & = I y_n^1 + J y_n^2 + K \sum_{k=2}^{p} \frac{J_1(k_F \sqrt{(x_n^1 - x_n^k)^2) + (y_n^1 - y_n^k)^2)}}{\sqrt{(x_n^1 - x_n^k)^2) + (y_n^1 - y_n^k)^2}} (y_n^1 - y_n^k) e^{-(k-1)/Me} \end{aligned} \tag{23}$$

with

$$\text{with } I = 1 + \frac{(1 - \frac{DT_F}{2m})}{(1 + \frac{DT_F}{2m})}, J = 1 - I, \text{ and } K = \frac{C h_0 T_F^2 k_F}{m(1 - \frac{DT_F}{2m})}$$

For $2 \leq i \leq p$, we get from (7) :

$$f_{2i-1}(x_n^1, y_n^1, \ldots, x_n^i, y_n^i \ldots, x_n^p, y_n^p) = x_n^{i-1} \tag{24}$$

$$f_{2i}(x_n^1, y_n^1, \ldots, x_n^i, y_n^i \ldots, x_n^p, y_n^p) = y_n^{i-1} \tag{25}$$

The $rot_\theta$ operator is a block diagonal square 2p matrix

$$rot_\theta = \begin{bmatrix} \cos\theta, -\sin\theta & & \\ \sin\theta, \cos\theta & & \\ & \ddots & \\ & & \cos\theta, -\sin\theta \\ & & \sin\theta, \cos\theta \end{bmatrix} \tag{26}$$

Eventually, a state vector taken at iteration 0 of a circular solution of dynamical system f is written from 13 as :

$$\vec{s} = \begin{bmatrix} R \\ 0 \\ \vdots \\ R\cos(1-i)\theta \\ R\sin(1-i)\theta \\ \vdots \\ R\cos(1-p)\theta \\ R\sin(1-p)\theta \end{bmatrix} \tag{27}$$

Equations 22, 23, 24, 25, 26, 27 allow to compute :

$$J_g(\vec{s}) = rot_{-\theta}.J_f(\vec{s})$$

with numerical tools. (Matlab symbolic package toolbox)

The spectral radius of $J_g(\vec{s})$ determines the stability of the self-orbit, after the eigenvalues 1 and $e^{\pm i\theta}$ corresponding respectively to the rotational and translationnal invariance of the dynamics are removed.

## Link to Supplementary Movies

Movie 1: https://youtu.be/qYF-6gM16Lo

Movie 2: https://youtu.be/ptjHvulshDs